\pdfoutput=1

\documentclass{PoS}

\usepackage{amsmath}
\usepackage{mathrsfs}          

\definecolor{rossoCP3}{cmyk}{0,.88,.77,.40}
\definecolor{verdeCP3}{rgb}{0.09765625, 0.57421875, 0.1015625}
\definecolor{bluCP3}{rgb}{0, 0.23, 0.67}
\definecolor{blUCLA}{rgb}{0, 0.4745, 0.647}

\newcommand{\eg}{e.g.~}

\newcommand{\Eq}[1]{Eq.~\eqref{#1}}
\newcommand{\Fig}[1]{Fig.~\ref{#1}}

\newcommand{\Lag}{\mathscr{L}}	

\newcommand{\beq}{\begin{equation}}
\newcommand{\eeq}{\end{equation}}

\newcommand{\ud}{\text{d}}
\newcommand{\bol}[1]{\mathbf{#1}}

\newcommand{\ER}{E_\text{R}}
\newcommand{\vmin}{v_\text{min}}
\newcommand{\tmax}{t_\text{max}}

\def\hhref#1{\href{http://arxiv.org/abs/#1}{#1}} 
\def\doi#1#2{\href{https://dx.doi.org/#1}{#2}} 

\title{Direct detection signals of dark matter with magnetic dipole moment}

\ShortTitle{Direct detection signals of dark matter with magnetic dipole moment}

\author{\speaker{Eugenio Del Nobile}
      \\
      Dipartimento di Fisica e Astronomia ``Galileo Galilei'', Universit\`a di Padova, \\ and INFN Sezione di Padova, Via Marzolo 8, 35131 Padova, Italy
      \\
      E-mail: \email{delnobile@cp3-origins.net}}

\abstract{
A neutral dark matter (DM) particle with a magnetic dipole moment has a very different direct detection phenomenology with respect to standard candidates. This is due to the peculiar functional form of the differential cross section for scattering with nuclei. Such a candidate could be a bound state of charged particles, as the neutron or an atom, or a fundamental particle coupled to heavier charged states, much like a Dirac neutrino. We analyze here the direct detection signals of DM with magnetic dipole moment, both the recoil rate and its modulation, and show that they are very different from those expected in standard scenarios. For this candidate, contrary to the common lore, the time of maximum signal depends on the recoil energy as well as on the target material. The observation of different modulations by experiments employing different targets would be a strong indication in favor of this type of DM particles.
}

\FullConference{The European Physical Society Conference on High Energy Physics\\
		 5-12 July, 2017
		 \\
		 Venice, Italy}

\begin{document}

\section{Introduction}
The analysis of direct dark matter (DM) detection data often relies on standard assumptions about the DM interactions with nucleons, namely the so-called spin-independent (SI) and spin-dependent (SD) interactions. However, other interactions are possible, each giving rise to a different signal. The rate spectrum experiments try to measure can thus be used to gain information about the DM properties. The same is also true for the time dependence of the rate, most notably its annual modulation due to Earth's rotation around the Sun. It is therefore important to have a clear view of what would be the signal produced not only by standard DM candidates, but also by less-standard yet motivated candidates.

A neutral DM particle with a magnetic dipole moment is interesting in this respect, since it has a very different direct detection phenomenology with respect to standard candidates, owing to the peculiar functional form of its differential scattering cross section with nuclei. It could arise as a bound state of charged particles, like the neutron or an atom, or be a fundamental particle which interacts with charged particles, much like a Dirac neutrino. Here we analyze the signals expected from this candidate, both the recoil rate and its modulation, and show that they are very different from those expected in the standard scenario. We also show that, contrary to the common lore, the time of maximum modulation depends on the recoil energy as well as on the target material. The observation of different modulations by experiments employing different targets would be a strong indication in favor of this type of DM particle. For more information see the \href{https://indico.cern.ch/event/466934/contributions/2587988/}{\textcolor{rossoCP3}{poster}}.

\section{Scattering rate}

Direct DM detection experiments try to measure the recoil energy $\ER$ a nucleus initially at rest in the detector acquires after scattering with a DM particle with initial velocity $\bol{v}$ in the detector's rest frame. The differential scattering rate on a nuclide $T$, neglecting factors inessential to this discussion, is
\beq\label{diffrate}
\frac{\ud R_T}{\ud \ER}(\ER, t) \sim \int_{v \geqslant \vmin(\ER)} \hspace{-3mm} v \, f(\bol{v}, t) \, \frac{\ud \sigma_T}{\ud \ER}(v, \ER) \, \ud^3 v \ ,
\eeq
where the $v \equiv |\bol{v}|$ factor comes from the DM flux. $f(\bol{v}, t)$ is the DM velocity distribution in Earth's frame, which depends on time due to Earth's motion around the Sun. $\vmin(\ER)$ is the minimum speed a DM particle must have to impart a fixed $\ER$ to a target nucleus, and its functional form is dictated by the scattering kinematics. In the case of elastic scattering treated here, there is a one-to-one correspondence between $\vmin$ and $\ER$, and these variables can be used interchangeably (see Ref.~\cite{DelNobile:2015rmp} for a discussion of the inelastic case).

\begin{figure}[t]
\centering
\hspace*{\fill}
\includegraphics[width=.4\textwidth, trim=0cm 13.6cm 22cm 0cm, clip=true]{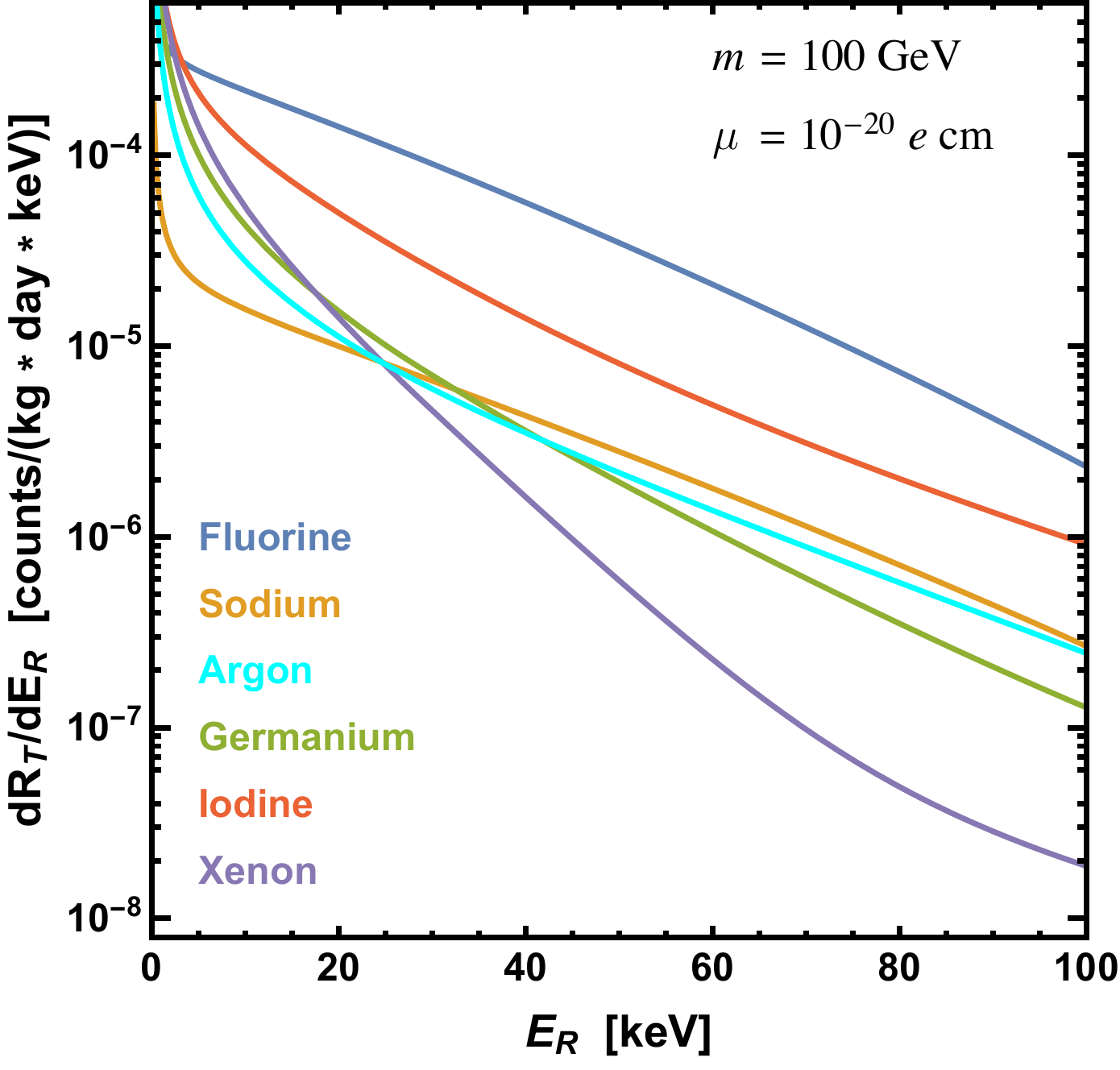}
\hfill
\includegraphics[width=.4\textwidth]{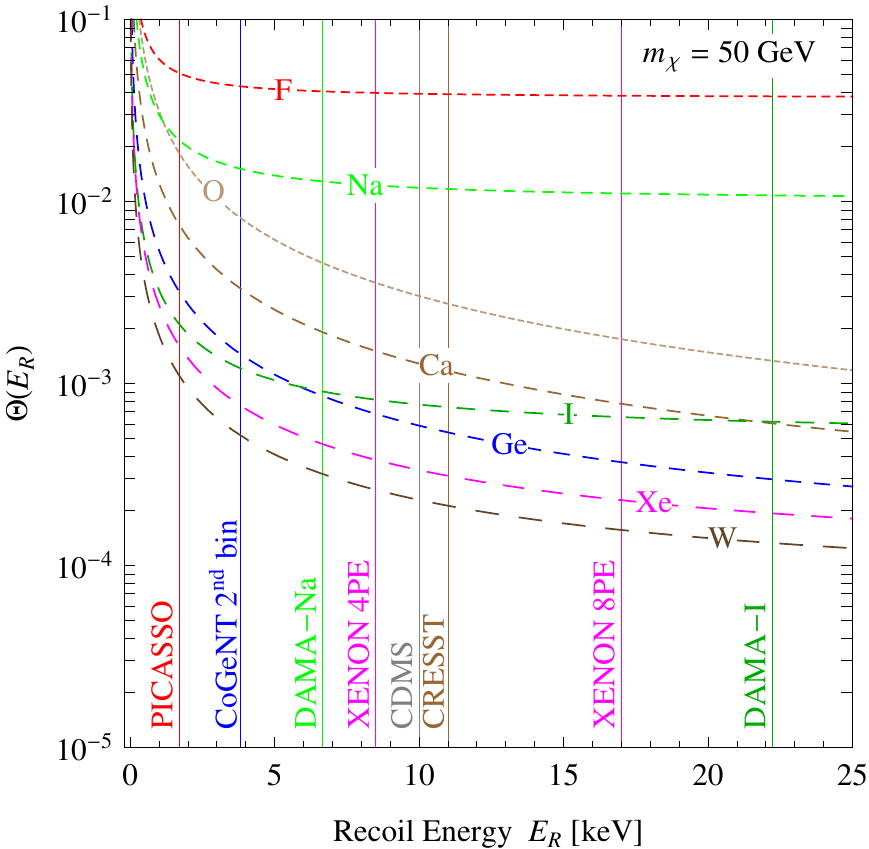}
\hspace*{\fill}
\caption{\label{fig:1}\em
Scattering rate for magnetic dipole DM (\textbf{left}) and its ratio with the rate for SI interactions (\textbf{right}).}
\end{figure}

The interaction Lagrangian of a Dirac fermion DM particle $\chi$ with magnetic moment $\mu$ is $\Lag = (\mu / 2) \, \bar{\chi} \sigma_{\mu \nu} \chi \, F^{\mu \nu}$. The differential cross section for elastic scattering off of a target nucleus with atomic number $Z$ and magnetic moment $\mu_T$ is (see \eg Refs.~\cite{DelNobile:2015rmp, DelNobile:2012tx, DelNobile:2015tza} and references therein)
\beq
\label{crosssect}
\frac{\ud \sigma_T}{\ud \ER}(v, \ER) = \underbrace{\left( \textcolor{red}{\frac{1}{\ER}} - \textcolor{verdeCP3}{\frac{\#}{v^2}} \right) \mu^2 \alpha Z^2 F^2_\text{C}(\ER)}_{\text{dipole-charge interaction}} \, + \! \underbrace{\textcolor{verdeCP3}{\frac{@}{v^2}} \mu^2 \mu_T^2 F^2_\text{M}(\ER)}_{\text{dipole-dipole interaction}} \! ,
\eeq
with $\#$ and $@$ two factors of no interest here. $F_\text{C}(\ER)$ and $F_\text{M}(\ER)$ are the Coulomb and magnetic form factors, respectively related to the SI and SD form factors. Two features are worth noting: \textit{(i)} the $1 / \ER$ behavior dominating the low-energy spectrum (see \Fig{fig:1}), and \textit{(ii)} the non-factorizable velocity dependence (colors in formulas mark different velocity dependences). In contrast, for the usual SI and SD interactions we have $\ud \sigma_T / \ud \ER \propto \textcolor{verdeCP3}{F^2(\ER) / v^2}$, with $F(\ER)$ the appropriate form factor.
Performing the velocity integral in \Eq{diffrate} we can write the differential scattering rate as
\begin{align}
\label{R&r}
\frac{\ud R_T}{\ud \ER}(\ER, t) = \textcolor{verdeCP3}{r_0(\ER, t)} + \textcolor{red}{r_1(\ER, t)}
&&
\text{with}
&&
r_n(\ER, t) \propto \eta_{n}(\vmin, t) \equiv \int_{v \geqslant \vmin} \hspace{-3mm} v^{2n} \, \frac{f(\bol{v}, t)}{v} \, \ud^3 v
\end{align}
for DM with magnetic moment, in contrast to $\ud R_T / \ud \ER \propto \textcolor{verdeCP3}{\eta_0}$ for SI/SD couplings. The peculiar recoil energy spectrum of this interaction, featuring the $1 / \ER$ divergence at low recoil energy, can be observed in the left panel of \Fig{fig:1}, showing the time-averaged rate for a $100$ GeV DM particle~\cite{DelNobile:2015rmp}. The right panel shows the ratio $\Theta(\ER)$ between the time-averaged rate for magnetic moment DM and that for DM with SI interactions (with arbitrary normalization), for a $50$ GeV DM particle~\cite{DelNobile:2012tx}. The target-dependent interplay of the dipole-charge and dipole-dipole terms in \Eq{crosssect}, which plays a major role for nuclei with large magnetic moments such as F (used \eg in PICO~\cite{Amole:2017dex}) and Na, I (used \eg in DAMA~\cite{Bernabei:2013xsa}), may be exploited to establish whether DM particles interact with nuclei through a magnetic dipole moment in case of detection.

\section{Time dependence}

\begin{figure}[t]
\centering
\hspace*{\fill}
\includegraphics[width=.4\textwidth, trim=0cm 1cm 13.8cm 1cm, clip=true]{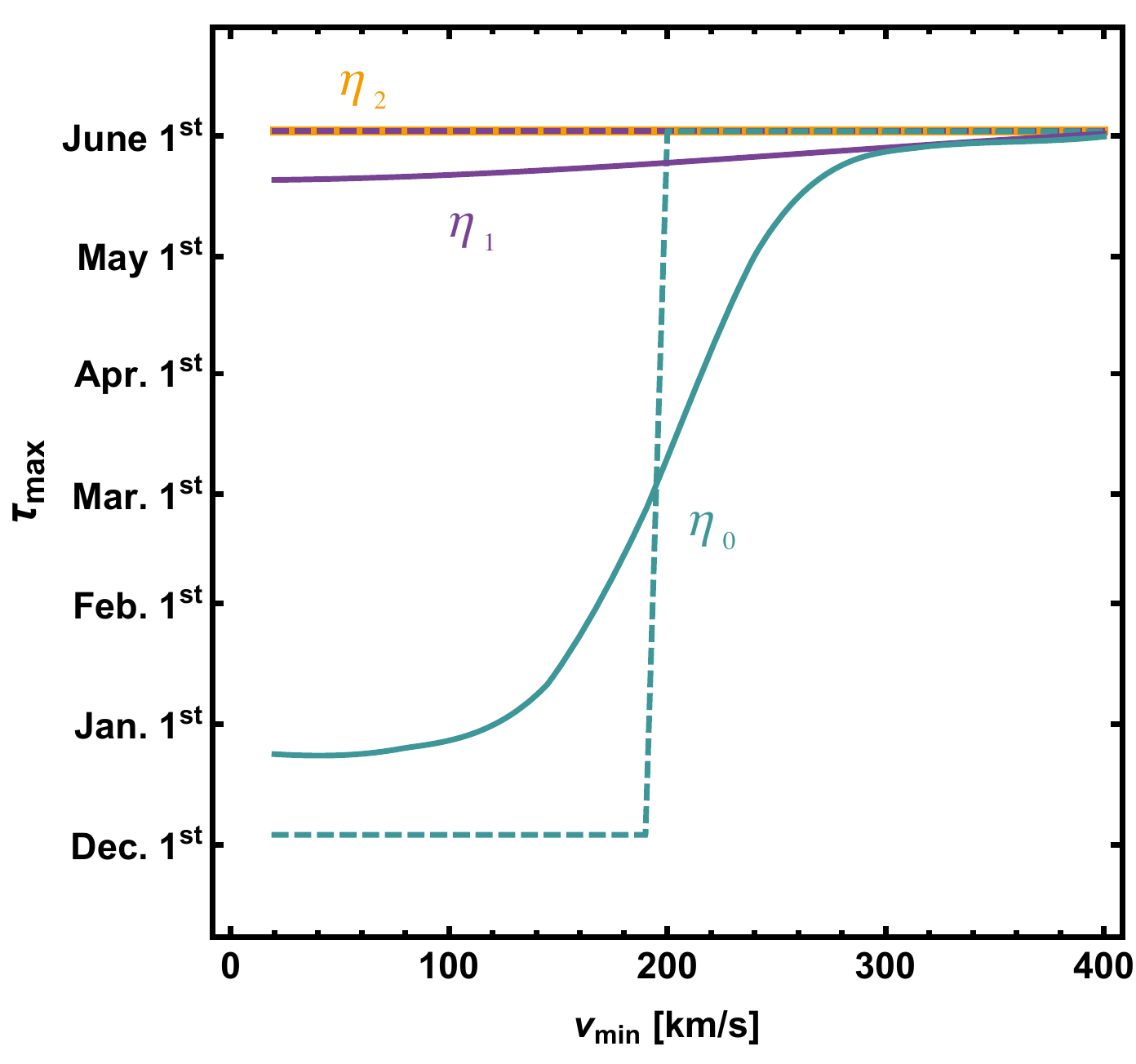}
\hfill
\includegraphics[width=.4\textwidth]{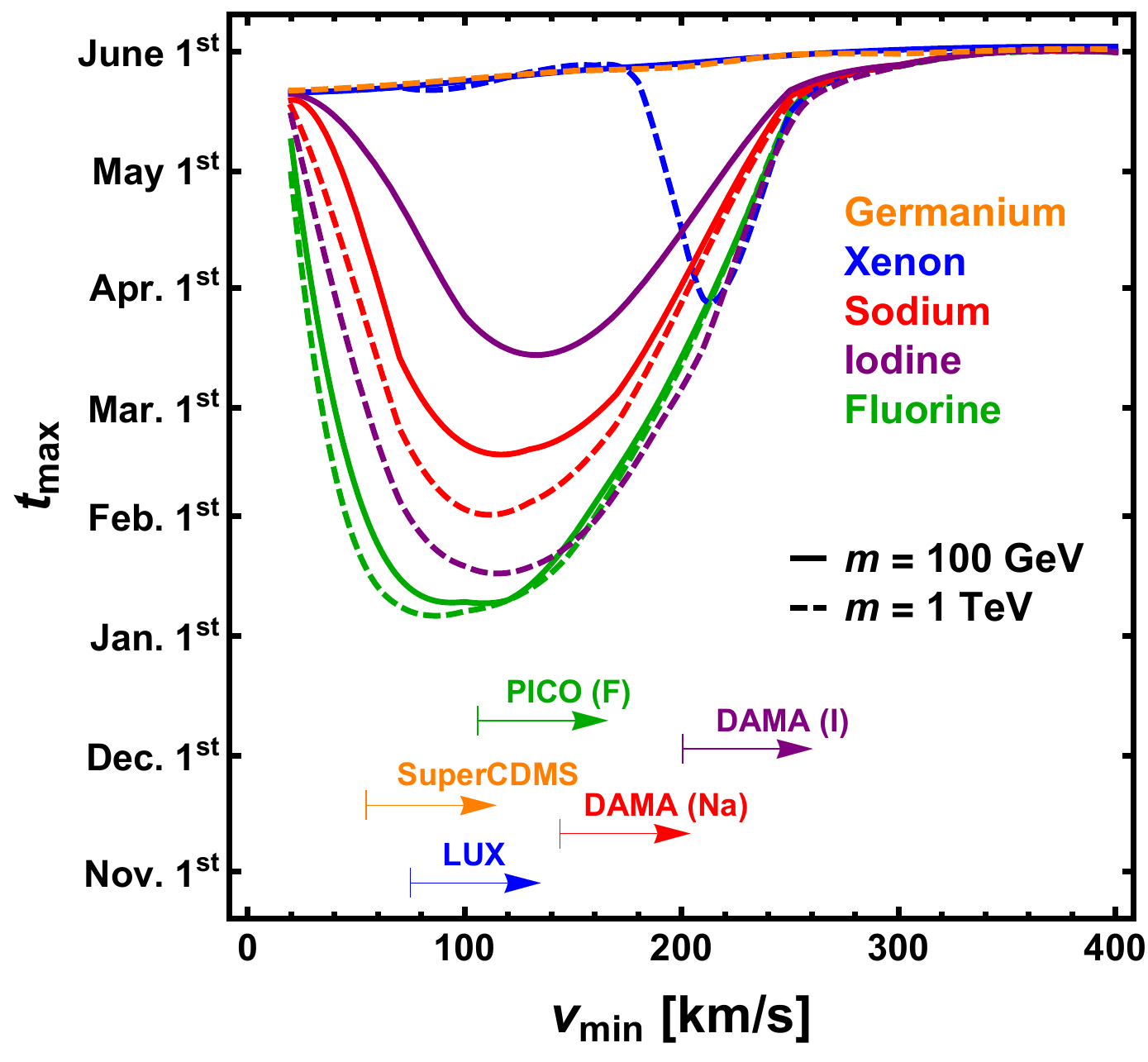}
\hspace*{\fill}
\caption{\label{fig:2}\em
\textbf{Left:} time of maximum of $\textcolor{verdeCP3}{\eta_0}$ and $\textcolor{red}{\eta_1}$ in the Standard Halo Model as a function of $\vmin$. \textbf{Right:} time of maximum of the differential scattering rate in \Eq{R&r} for magnetic dipole DM.}
\end{figure}

In general, the scattering rate depends on time through the velocity integral in \Eq{diffrate}. The left panel of \Fig{fig:2} displays the time of maximum of $\textcolor{verdeCP3}{\eta_0}$ and $\textcolor{red}{\eta_1}$ (see \Eq{R&r}) in the Standard Halo Model (solid lines take into account the effect of the gravitational potential of the Sun on the DM velocity distribution, dashed lines ignore it). If the target and velocity dependences can be factored in the differential rate, as happens \eg for the SI/SD interactions where $\ud R_T / \ud \ER \propto \textcolor{verdeCP3}{\eta_0}$, the rate depends on time through a single velocity integral, the same function for all targets. Quantities such as the time of maximum rate $\tmax$ (as well as the fractional amplitudes and phases in a Fourier expansion of the signal) are then independent of the target material when expressed as functions of $\vmin$ (as opposite to $\ER$).

For DM with a magnetic dipole, this feature is spoiled by the target-dependent interplay of terms with different velocity dependences ($\textcolor{verdeCP3}{1 / v^2}$ and $\textcolor{red}{1 / \ER}$ in \Eq{crosssect}, or $\textcolor{verdeCP3}{r_0}$ and $\textcolor{red}{r_1}$ in \Eq{R&r}). $\textcolor{red}{r_1}$ always dominates at low recoil energies thanks to the $1 / \ER$ enhancement, while $\textcolor{verdeCP3}{r_0}$ always dominates at large enough $\ER$~\cite{DelNobile:2015tza}. This implies that the time dependence of the rate is dictated by $\textcolor{red}{\eta_1}$ at low $\vmin$ and by $\textcolor{verdeCP3}{\eta_0}$ at large $\vmin$. This behavior is apparent in the right panel of \Fig{fig:2}, showing $\tmax$ as a function of $\vmin$, with the small arrows indicating some experimental thresholds~\cite{DelNobile:2015tza}. As one can see, $\textcolor{verdeCP3}{r_0}$ gets to dominate the rate at low enough $\vmin$ to produce a detectable feature in $\tmax$ only in target nuclei with a sizeable magnetic moment, and/or for large enough DM masses.

The plots above are for a fully differential (unintegrated) distribution, but a putative signal will realistically need to be binned. As discussed in Ref.~\cite{DelNobile:2015rmp}, it should be possible to distinguish DM with magnetic dipole interactions from DM with SI/SD interactions relying on the time dependence of the rate alone, by integrating the signal in $1$ keV bins. Integration from a certain threshold up, as performed \eg in PICO, will lead to indistinguishable signals. The observation of a different time of maximum signal by experiments employing different targets would be a strong indication in favor of this type of DM particles.

\end{document}